# Deep Learning Based Signal Enhancement of Low-Resolution Accelerometer for Fall Detection Systems

Kai-Chun Liu, Kuo-Hsuan Hung, Chia-Yeh Hsieh, Hsiang-Yun Huang, Chia-Tai Chan, and Yu Tsao

*Abstract*— In the last two decades, fall detection (FD) systems have been developed as a popular assistive technology. To support long-term FD services, various power-saving strategies have been implemented. Among them, a reduced sampling rate is a common approach for an energy-efficient system in the real world. However, the performance of FD systems is diminished owing to low-resolution (LR) accelerometer signals. To improve the detection accuracy with LR accelerometer signals, several technical challenges must be considered, including mismatch of effective features and the degradation effects. In this work, a deep-learning-based accelerometer signal enhancement (ASE) model is proposed as a front-end processor to help typical LR-FD systems achieve better detection performance. The proposed ASE model based on a deep denoising convolutional autoencoder architecture reconstructs high-resolution (HR) signals from the LR signals by learning the relationship between the LR and HR signals. The results show that the FD system using support vector machine and the proposed ASE model at an extremely low sampling rate (sampling rate < 2 Hz) achieved 97.34% and 90.52% accuracies in the SisFall and FallAllD datasets, respectively, while those without ASE models only achieved 95.92% and 87.47% accuracies in the SisFall and FallAllD datasets, respectively. The results also demonstrate that the proposed ASE mode can be suitably combined with deep-learning-based FD systems.

## I. INTRODUCTION

F ALLS are a leading cause of major injuries and deaths among the aging population. The world health organization [1] reported that more than 30% of the elderly, aged over 64 years, fall at least once a year. Approximately half of them were unable to get back up without assistance due to fall-related injuries or a lack of physical fitness and strength [2]. A previous study has shown that 50% of the elderly population died within 6 months of the fall event because they were lying on the floor for more than an hour [3].

This work was supported in part by grants from the Ministry of Science and Technology, Taiwan, under Grant MOST 109-2221-E-001-022 and Grant 108-2628-E-001 -002 -MY3, 109-2634-F-008 -006, and Academia Sinica, under Grant AS-CDA-106-M04 and AS-GC-109-05.
(Corresponding author: Yu Tsao, e-mail: yu.tsao@citi.sinica.edu.tw).

K.-C. Liu, K.-H. Hung, and Y. Tsao are with the Research Center for Information Technology Innovation, Academia Sinica, Taipei 11529, Taiwan.

Y. Tsao is also a jointly appointed professor of the Department of Electrical Engineering, Chung Yuan Christian University, Taoyuan 32023, Taiwan

C.-Y. Hsieh, H.-Y. Huang and C.-T. Chan are with the Department of Biomedical Engineering, National Yang Ming Chiao Tung University, Taipei 11221, Taiwan.

C.-Y. Hsieh is with Bachelor's Program in Medical Informatics and Innovative Applications, Fu Jen Catholic University, New Taipei City 24205, Taiwan.

In the past two decades, automatic fall detection (FD) systems have been developed as an assistive technology [4]. The main goal of FD systems is to automatically detect critical fall events and immediately alert medical professionals or caregivers [4-6]. Moreover, these systems can relieve the psychological stress affecting elders and caregivers [7].

Owing to advancements in information communication technologies and body sensor networks, various sensor technologies have been used in FD systems including inertial sensors [6], depth-cameras [8], microphones [9], pressure sensors [10], and thermal sensors [11]. In particular, accelerometers are the most common sensors for FD, which capture body movements and are sensitive to posture changes. FD systems using accelerometers have advantages such as compactness, low cost, effectiveness, unobtrusiveness, and high mobility [12].

Wearable sensors must be placed on the body and function as long as possible to enable long-term FD services. This demand leads to challenges in the design and development of systems, including reliability, security, usability, and sustainability [13]. For instance, the number of batteries would impact the size and comfortability of sensors [14]. Furthermore, recharging or replacing batteries frequently may decrease its usability [15] and the user's acceptance of the FD system [16]. Therefore, several energy-efficient FD systems that focus on reducing power consumption and extending battery life have been developed for long-term FD services.

In a typical FD system, measurement devices (e.g., wrist-band and smartwatches) send the collected data wirelessly to a processing unit (e.g., smartphones, laptops, and workstations) for identifying a fall event. Factors such as sampling rate, feature extraction and selection, communication protocol, detector design, and utilization of low-power electronic components [17-22] influence the power consumption in FD systems. Most studies adjust or optimize the sampling rates for power-efficient FD systems. This is because, at high sampling rates, up to 90% of the power consumption of wearable health monitoring systems is owing to data sampling [23, 24]. Moreover, the lesser the data collected, the lesser the data transfer volume. This could reduce the power consumption in FD systems.

Various machine learning (ML) techniques [12] have been applied to improve the detection performance of the typical FD systems [25] as they mainly rely on human knowledge to build rule-based FD classifiers, including k-nearest-neighbors and support vector machines. Furthermore, the advanced deep learning (DL) techniques have shown superiority in the detection ability, such as convolutional neural networks and long short-term memory [26]. However, previous works [19, 20] have reported the performance of ML-based FD systems is sensitive to the sampling rates. In particular, the detection accuracy of the FD system decreases significantly when the sampling rate is less than 10 Hz. This is because low sampling rates lead to low-resolution (LR) accelerometer signals for typical FD systems. The negative effects of LR signals would hinder the reliability of FD systems.

To tackle the technical issue of LR signals, Wang *et al.* [17, 18] proposed a power-saving framework that used a trigger module to switch from a low-power mode with a sampling rate of 6 Hz to the measurement mode at 50 Hz when the possible fall event is triggered. Their approaches can provide an estimated battery life of 664.9 days [17] and 1,125 days [18] for FD systems using both accelerometer and barometric pressure and a single accelerometer, respectively. However, different from the previous works [17, 18], the present study focused on directly enhancing the detection performance of LR-FD systems without an additional trigger module.

Improving the detection accuracy of typical FD systems using LR accelerometer signals requires a consideration of two technical challenges. The first challenge is the *mismatch of effective features*. LR accelerometer signals lead to a loss in fine-grained movement information. The feasible features used in detection systems that are trained using high-resolution (HR) signals cannot perform well when LR signals are processed. In other words, additional efforts are required to explore the *effective features* of LR-FD systems. The second one is the *degradation effects*. The critical fall characteristics including free-fall, impact, vibration, and recovery are degraded in low-quality signals. Such degradation makes it difficult to tackle classification problems (e.g., inter-class ambiguity and intra-class variability).

In this study, we aim to propose a DL-based accelerometer signal enhancement (ASE) model to help typical FD systems tackle the aforementioned challenges and achieve better detection performance in LR-FD systems. The proposed DL-based ASE model is applied as a front-end processor to reconstruct HR signals from LR signals. The reconstructed HR signals are then fed to the FD system. The classical FD systems can yield higher detection accuracy with reconstructed HR signals.

The main contributions of this work are as follows:

- A DL-based ASE model, based on a deep denoising convolutional autoencoder architecture, is proposed to reconstruct HR accelerometer signals from LR ones.
- We comprehensively analyze the DL-based ASE model on wearable FD systems and typical FD systems to measure the detection performance, which varies with different sampling rates.
- Two emulated public FD datasets were employed to validate the effectiveness of the proposed ASE model for achieving a higher detection accuracy.

The remainder of this paper is organized as follows. Section II introduces the selected open datasets and their experimental protocols. In Section III, we describe the design principles and mechanisms of the proposed ASE model for FD systems. The results of the experiment with the proposed FD system are presented in Section IV. The comprehensive performance analysis of the ASE model for FD systems, its limitations, and future works are discussed in Section V. Finally, we conclude this study in Section VI.

## II. OPEN DATASETS

Currently, open datasets for wearable FD including SisFall [27], FallAllD [26], UMAFall [28], and UPFall [29] are available. In this study, SisFall and FallAllD datasets are used to validate the effectiveness of the proposed DL-based ASE model on the LR-FD. Compared to other datasets, these two datasets with diverse fall types and activities of daily living (ADL) are more challenging and closer to the real-world situation. In addition, these datasets use accelerometers with a sampling rate of at least 200 Hz. This may contrast the effects of the proposed model with different resolution accelerometer signals from HR signals to LR signals.

### A. FallAllD dataset

The FallAllD dataset proposed by Saleh and Jeannes [26] was used in this study. An inertial measurement unit (IMU) was placed on the neck, chest, and waist, respectively, to measure the movement during the experiment. Each inertial sensor unit includes a tri-axial accelerometer (sampling rate: 238 Hz, range: ± 8 g), tri-axial gyroscope (sampling rate: 238 Hz, angular rate: ± 2000 dps), tri-axial magnetometer (sampling rate: 80 Hz, range: ± 4 G), and a barometer (sampling rate: 10 Hz). A previous study [30] showed that the FD system using a waist-worn accelerometer is better than those in other positions. Therefore, we focus on the collected accelerometer data from the waist.

The experimental data were collected from 15 healthy young subjects (8 males and 7 females). Their average age, height, and weight were 32 years, 171 cm, and 67 kg, respectively. We only used the data from 10 subjects that performed falls and ADL. In total, this dataset has 1053 ADL and 423 fall trials.

### B. SisFall dataset

The SisFall dataset [27] involves two age groups: 15 elder healthy subjects and 23 young healthy subjects. This study employs the raw data collected from 21 young subjects (10 males, 11 females, age $25.0 \pm 8.6$ years, height $165.7 \pm 9.3$ cm, weight $57.7 \pm 15.5$ kg) for the experiment. A group of elder people was not considered in this study because they were not made to fall in these experiments. In addition, two young adults were excluded owing to their incomplete ADL trials. Totally, this dataset has 1575 fall and 1659 ADL trials. An IMU (Shimmer, Ireland) placed on the waist captured the motion data at a sampling rate of 200 Hz. The proposed model only explores the data collected from accelerometers, while the sensor consists of accelerometers and gyroscopes.

### C. Downsampling

The downsampling approach is employed to obtain LR signals from HR signals for exploring the effects of different resolutions on the FD performance. The LR signals $S^{LR}$ are gathered by applying downsampling approaches to the HR signals $S^{HR} = \{s_j | j = 1, 2, \ldots, n_{HR}\}$ by an integer factor $n$, which is defined as follows:

$$S^{LR} = \{s_k | k = 1 + 2^\alpha \times n\}, \tag{1}$$

where it keeps the first sample from every $2^\alpha$ sample, $0 \le n \le \frac{n_{HR}-1}{2^\alpha}$, and $n_{HR}$ is the total number of samples in $S^{HR}$. The corresponding sampling rate of the LR signals is $R/2^\alpha$ Hz, where the original sampling rate of the HR accelerometer signals is $R$ Hz. In this study, an integer factor ranging from $\alpha = 1,2,...,7$ is applied to the SisFall and FallAllD datasets.

## III. METHODOLOGY

This study develops an FD system with a DL-based ASE model as a front-end processor for FD using LR accelerometer signals. The overall system consists of the pre-processing stages, the ASE model, and the FD system. Initially, a series of signal pre-processing stage is applied to the raw LR accelerometer signals $S^{LR}$, involving impact-defined window and min-max normalization. The network architecture of the proposed deep ASE model involves feed-forward convolutional neural networks and deep neural networks. The main goal of the ASE model is to enhance the LR accelerometer signals and generate HR signals. Finally, the classical ML-based classifier is applied to the enhanced signals to classify the fall and ADL. The framework of the proposed DL-based ASE model for FD system is introduced in Fig. 1.

### A. Pre-processing

#### 1) Impact-defined window

The impact-defined window approach focuses on covering the critical fall patterns (pre-impact, impact and post-impact) in a window based on an impact point that is determined as the moment when the human body hits the first determines the highest peak as the impact point within a searching area. Then, the fixed window sizes are applied to cover the patterns before and after the impact point.

Given that the input accelerometer signals of a trial are defined as $S = \{s_j | j = 1,2,...,n_s\}$, where $n_s$ is the total number of samples in a trial. The sample point involves the three-dimensional acceleration $s_j = \{a_{x_j}, a_{y_j}, a_{z_j}\}$. First, the data sample with the maximum $Norm_{xyz}$ is considered as the impact sample point $s_I$ of the trial, where $Norm_{xyz}$ of $s_j$ is calculated by (2),

$$Norm_{xyz}(s_j) = \sqrt{a_{x_j}^2 + a_{y_j}^2 + a_{z_j}^2}, \tag{2}$$

Next, based on the $s_I$, forward and backward sub-windows are determined as $W_f = \{s_{I+1},...,s_{I+WS_f-1}, s_{I+WS_f}\}$ and $W_b = \{s_{I-WS_b}, s_{I-WS_b+1},...,s_{I-1}\}$, respectively. Here, $WS_f$ and $WS_b$ are the window sizes of $W_f$ and $W_b$, respectively. Details of the impact-defined window approach are introduced in [6, 30, 31].

$WS_f$ and $WS_b$ are determined as 2s and 1.44s, respectively for the SisFall dataset, and 2s and 1.23s, respectively for the FallAllD dataset. A previous study has shown that typical FD systems with such window sizes can achieve the best system

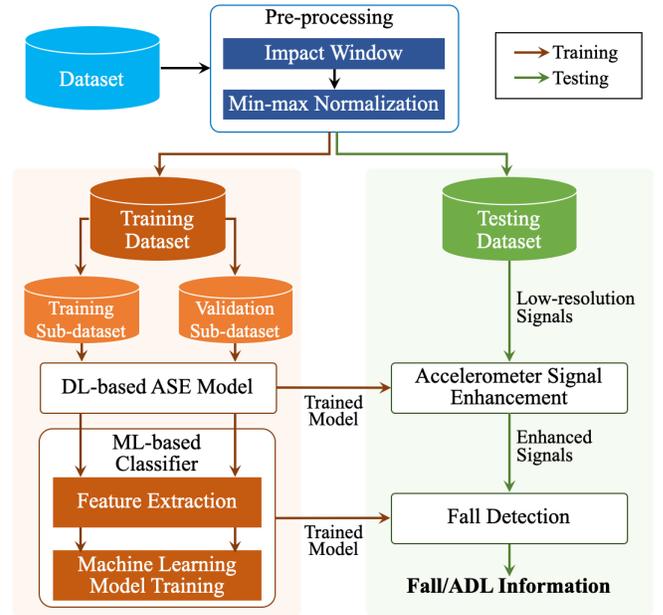

Fig. 1 The framework of the proposed DL-based ASE for FD systems.

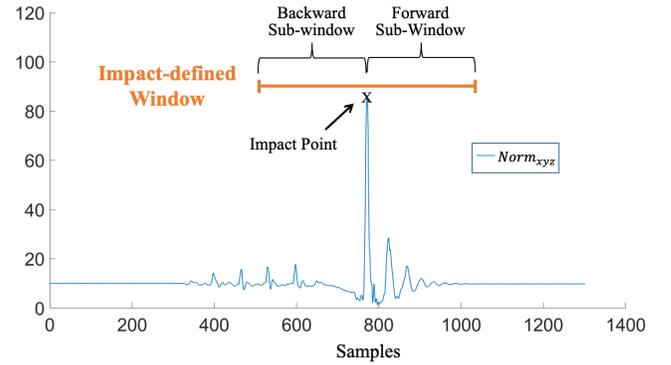

Fig. 2 An illustration of the impact-defined window on a fall trial [31].

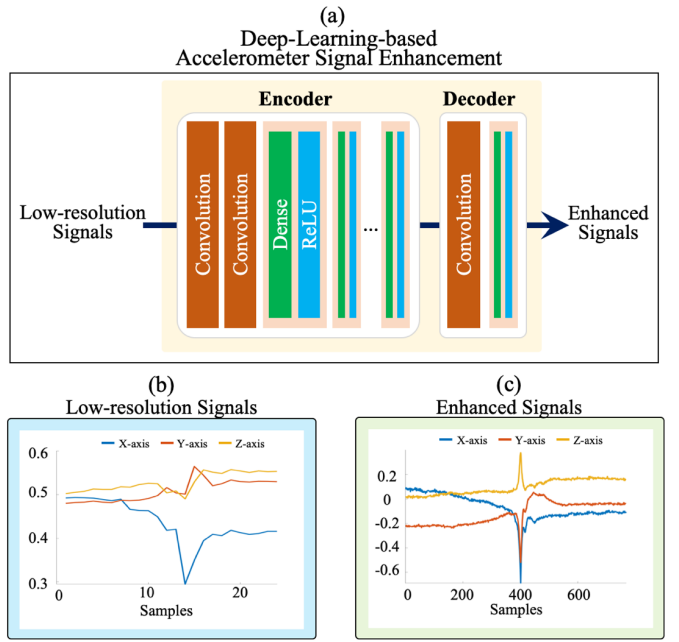

Fig. 3 (a) The architecture of the proposed DL-based accelerometer signal enhancement model. The example of the (b) low-resolution signals and (c) enhanced signals.

TABLE I
THE IMPLEMENTATION DETAIL OF THE PROPOSED ASE MODEL FOR DIFFERENT SAMPLING RATES.

| Sampling Rate (Hz) | | Encoder Convolution layer | | | | Encoder Dense layer | | Decoder Convolution layer | | | Decoder Dense layer | |
| --- | --- | --- | --- | --- | --- | --- | --- | --- | --- | --- | --- | --- |
| SisFall | FallAllD | Input size | Layers | Filter size | Output channel | Layers | Hidden units | Layers | Filter size | Output channel | Layers | Hidden units |
| 200 | 238.00 | 768 | 2 | 3×3 | 40 | 5 | [768,768,768, 768,768] | 1 | 3×3 | 1 | 1 | 768 |
| 100 | 119.00 | 384 | 2 | 3×3 | 35 | 6 | [384,384,384, 384,768,768] | 1 | 3×3 | 1 | 1 | 768 |
| 50 | 59.50 | 192 | 2 | 3×3 | 30 | 6 | [192,192,192, 384,768,768] | 1 | 3×3 | 1 | 1 | 768 |
| 25 | 29.75 | 96 | 2 | 3×3 | 25 | 8 | [96,96,96,96,192, 384,768,768] | 1 | 3×3 | 1 | 1 | 768 |
| 12.5 | 14.88 | 48 | 2 | 3×3 | 20 | 8 | [48,48,48,96,192, 384,768,768] | 1 | 3×3 | 1 | 1 | 768 |
| 6.25 | 7.44 | 24 | 2 | 3×3 | 15 | 8 | [24,24,48,96,192, 384,768,768] | 1 | 3×3 | 1 | 1 | 768 |
| 3.13 | 3.72 | 12 | 2 | 3×3 | 10 | 8 | [12,24,48,96,192, 384,768,768] | 1 | 3×3 | 1 | 1 | 768 |
| 1.56 | 1.86 | 6 | 2 | 3×3 | 5 | 8 | [12,24,48,96,192, 384,768,768] | 1 | 3×3 | 1 | 1 | 768 |

TABLE II
LIST OF THE EXTRACTED FEATURES [31]

| No. | Features |
| --- | --- |
| $f_1$-$f_6$ | Mean of $a_{x_j}^E$, $a_{y_j}^E$, $a_{z_j}^E$, $a_{norm_j}^E$, $a_{verti_j}^E$, $a_{horti_j}^E$ |
| $f_7$-$f_{12}$ | Standard Deviation of $a_{x_j}^E$, $a_{y_j}^E$, $a_{z_j}^E$, $a_{norm_j}^E$, $a_{verti_j}^E$, $a_{horti_j}^E$ |
| $f_{13}$-$f_{18}$ | Variance of $a_{x_j}^E$, $a_{y_j}^E$, $a_{z_j}^E$, $a_{norm_j}^E$, $a_{verti_j}^E$, $a_{horti_j}^E$ |
| $f_{19}$-$f_{24}$ | Maximum of $a_{x_j}^E$, $a_{y_j}^E$, $a_{z_j}^E$, $a_{norm_j}^E$, $a_{verti_j}^E$, $a_{horti_j}^E$ |
| $f_{25}$-$f_{30}$ | Minimum of $a_{x_j}^E$, $a_{y_j}^E$, $a_{z_j}^E$, $a_{norm_j}^E$, $a_{verti_j}^E$, $a_{horti_j}^E$ |
| $f_{31}$-$f_{36}$ | Range of $a_{x_j}^E$, $a_{y_j}^E$, $a_{z_j}^E$, $a_{norm_j}^E$, $a_{verti_j}^E$, $a_{horti_j}^E$ |
| $f_{37}$-$f_{42}$ | Kurtosis of $a_{x_j}^E$, $a_{y_j}^E$, $a_{z_j}^E$, $a_{norm_j}^E$, $a_{verti_j}^E$, $a_{horti_j}^E$ |
| $f_{42}$-$f_{48}$ | Skewness of $a_{x_j}^E$, $a_{y_j}^E$, $a_{z_j}^E$, $a_{norm_j}^E$, $a_{verti_j}^E$, $a_{horti_j}^E$ |
| $f_{49}$-$f_{51}$ | Correlation coefficient between each pair of $a_{x_j}^E$, $a_{z_j}^E$ |
| $f_{52}$-$f_{54}$ | Correlation coefficient between each pair $a_{norm_j}^E$, $a_{verti_j}^E$, $a_{horti_j}^E$ |

Note. $a_{x_j}^E$, $a_{y_j}^E$, $a_{z_j}^E$, $a_{norm_j}^E$, $a_{verti_j}^E$, $a_{horti_j}^E \in S_j^E$, where $j = 1,2,...,N$

performance [31]. In this work, the searching area is determined as the signals of the trial. Therefore, the impact-defined window approach applied to each trial can obtain an impact-defined window. Fig. 2 illustrates the impact-defined window on a trial.

### 2) Min-max normalization

After the impact-defined window, a min-max normalization is utilized to process the data. This approach can reduce scaling effects on the deep ASE model during the training phase. Given a data sequence $S = \{s_j | i = 1,2,...,n_S\}$, $s_j$ can be normalized to the range [0,1] using min-max normalization, as shown in Equation (3):

$$s_j^{nom} = \frac{s_j - s_{min}}{s_{max} - s_{min}}, \tag{3}$$

where $s_{max}$ and $s_{min}$ are the maximum and minimum values of $S$, respectively.

### B. DL-based ASE

#### 1) Network architecture of the ASE model

The architecture of the proposed ASE model for FD systems is shown in Fig. 3. The over-model is a deep denoising autoencoder architecture, which consists of an encoder and decoder that contain convolution layers and dense layers. First, the encoder convolution layer (ECL) captures movement features from a tri-axial accelerometer signal. Each ECL includes two convolution layers, consisting $5 \times (8 - log_2(\frac{O}{D}))$ filters with a filter size of 3×3 and the stride of 1, where $O$ and

$D$ are the sampling rates of the HR and LR signals, respectively. Then, the encoder dense layer reconstructs the HR signals from the LR signals. The encoder dense layer consists of 5 to 8 layers. Finally, the decoder adjusts the encoder output to be similar to the original data. All decoder convolution layers in the ASE model have a filter size of 3×3 and a stride of 1. All dense layers employ rectified linear units (ReLU) as the activation function for the hidden layers. The convolution layers do not involve activation functions. The details of the ASE models at different sampling rates are shown in Table IV.

#### 2) Training strategy

Given that the preprocessed HR and LR accelerometer signals are defined as $S^{HR} = \{s_1^{HR}, s_2^{HR}, ..., s_N^{HR}\}$ and $S^{LR} = \{s_1^{LR}, s_2^{LR}, ..., s_M^{LR}\}$, respectively, where $N \geq 2^\alpha \times M$, we use LR-HR signal pairs to train the ASE model. The enhanced accelerometer signals $S^E = \{s_1^E, s_2^E, ..., s_N^E\}$ are represented as:

$$S^E = f_{decoder}(f_{encoder}(S^{LR})), \tag{4}$$

where $f_{encoder}$ and $f_{decoder}$ denote the encoder and decoder functions, respectively. The mean absolute error (MAE) is used as the objective function, and the loss function is formulated as:

$$loss = \frac{1}{N}\sum_{j=1}^{N}\left|S_j^{HR} - S_j^E\right|, \tag{5}$$

where $S_j^{HR}$ and $S_j^E$ denotes the HR and the enhanced signals, respectively, and $N$ denotes the number of training sets. The ASE model is trained with the training data that is randomly divided into a training and a validation sub-dataset at a ratio 9:1. In the encoder, the layer numbers are set according to the sampling rate. This is because the lower sampling rate offers less information than higher sampling rate. Therefore, fewer filter numbers are sufficient for feature extraction through the ASE model. Note that the pooling layer and dropout are not used in the model.

### C. FD system

In this study, feature extraction and ML models are applied to the enhanced signals for fall event detection. The feasibility of these processes have been validated in previous works [6, 30-32].

TABLE III
THE BEST PERFORMANCE USING TYPICAL WEARABLE FD SYSTEMS WITH
AND WITHOUT THE ASE MODEL IN SISFALL DATASET (%).

|  | kNN_original | kNN_enhanced | SVM_original | SVM_enhanced |
|---|---|---|---|---|
| Accuracy | 98.70 | 99.07 | 98.95 | **99.41** |
| Sensitivity | 98.22 | 99.11 | 99.24 | **99.56** |
| Specificity | 99.22 | 99.04 | 98.79 | **99.34** |
| Precision | 99.17 | 98.99 | 98.73 | **99.30** |

TABLE IV
THE BEST PERFORMANCE USING TYPICAL WEARABLE FD SYSTEMS WITH
AND WITHOUT THE ASE MODEL IN THE FALLALLD DATASET (%).

|  | kNN_original | kNN_enhanced | SVM_original | SVM_enhanced |
|---|---|---|---|---|
| Accuracy | 94.11 | 93.97 | 94.58 | **95.60** |
| Sensitivity | 87.00 | **90.31** | 88.89 | 90.31 |
| Specificity | 96.96 | 95.54 | 96.96 | **97.72** |
| Precision | 92.00 | 89.02 | 92.12 | **94.09** |

*1) Feature extraction*

Feature extraction aims to obtain effective parameters that can be used to classify ADL and fall events. The FD system utilizes eight one-dimensional and one three-dimensional statistical features to enhance the accelerometer signals including mean, standard deviation, variance, maximum, minimum, range, kurtosis, skewness, and correlation coefficient. Previous studies have shown the effectiveness of these features for typical FD systems [6, 30-32]. Firstly, we calculate three important values from the enhanced accelerometer signals, including Euclidean norm acceleration $a_{norm}$, Euclidean norm on vertical and horizontal planes $a_{verti}$, $a_{horti}$. Then, the feature extraction is applied to a sequence of the enhanced accelerometer signals $S^E = \{s_1^E, s_2^E, ..., s_N^E\}$ to obtain a set of features $F = \{f_1, f_2, ... f_k\}$, where $k$ is the number of extracted features. Any $s_j^E$ is denoted as $\left\{a_{x_j}^E, a_{y_j}^E, a_{z_j}^E, a_{norm_j}^E, a_{verti_j}^E, a_{horti_j}^E\right\}$ involving three-dimensional accelerometer signals and three important values, $1 \leq j \leq N$, and $s_j^E \in S^E$. 54 features gathered from $S^E$ are applied to ML models. The utilized features are listed in TABLE II.

*2) ML-based classifier*

Two classical ML classifier were utilized to analyze the effects of the DL-based ASE model on the LR-FD systems: support vector machine (SVM) and $k$-nearest neighbor (kNN). Previous studies have shown that the selected ML classifiers are relatively more reliable, as compared to other common ML classifier (e.g., naïve Bayes and decision tree), for FD systems [6, 30-32]. Brief introductions of these classification algorithms are as follows:

- **SVM**: Generally, the SVM model distinguishes target classes with maximum margins based on the hyperplane optimization. The input data are first mapped onto a new dimensional space. Then, the decision boundary is determined by finding the largest possible margin between the points of the different classes. Finally, the SVM model classifies the testing data according to the decision boundary. In this study, a radial basis function (RBF) kernel is employed for the SVM. The classifier is optimized with a Bayesian optimizer, where both kernel scale (gamma) and box constraint are set to 1.

- **kNN**: Known also as a lazy classifier, is a typical instance-based classification model. This classification mainly relies on majority voting among the $k$-nearest training samples to the testing sample. The Euclidean distance is taken into account to compute the weight of the neighbors. The previous study [31] has shown that the classical FD system has the best detection accuracy with $k = 3$ with the Euclidean distance. Therefore, 3NN and the Euclidean distance are employed to classify fall/ADL in this study.

*D. Performance evaluation*

Leave-one-subject-out (LOSO) cross-validation approach is introduced to validate the effectiveness of the proposed ASE model on FD systems. This approach can evaluate performance on unseen subjects and lead to more reliable models. LOSO approach keeps the data from one of the subjects as the testing dataset; the data consisting of other subjects is considered as the training dataset. Then, repeating $k$ times until $k$ subjects have been used as the testing dataset, where $k$ is the total number of the subjects. Finally, the average of the $k$ folds performance is outputted as the system performance.

Four evaluation metrics, accuracy (ACC), sensitivity (SEN), specificity (SPE), and precision (PRE), were introduced to measure the detection performance. These metrics are defined in Equations (15)–(18):

$$ACC = \frac{TP+TN}{TP+FP+TN+FN}, \tag{15}$$

$$SEN = \frac{TP}{TP+FN}, \tag{16}$$

$$SPE = \frac{TN}{TN+FP}, \tag{17}$$

$$PRE = \frac{TP}{TP+FP}, \tag{18}$$

where true positive (TP), true negative (TN), false positive (FP), and false negative (FN) are cases where the labeled fall signals are recognized as a fall, the labeled ADL signals are recognized as an ADL, the labeled ADL signals are recognized as a fall, and the labeled fall signals are recognized as an ADL, respectively.

The ASE model was implemented through PyTorch 1.3.1, running on a workstation with 64 bit Ubuntu 18.04.4, Intel(R) Xeon(R) Gold 6152 CPU @ 2.10GHz, and trained and tested using the Nvidia GTX 2080Ti with 11 GB dedicated memory. The pre-processing and FD processes are realized using the Statistics and Machine Learning Toolbox in the MATLAB 2016 environment.

## IV. EXPERIMENTAL RESULTS

*A. Performance analysis on the ASE-based FD systems*

In this study, typical FD systems without the proposed ASE model are used as a baseline. During the training phase, the selected parameters of FD systems are fixed for both baseline and ASE-based FD systems. The detection performance of the FD systems using kNN and SVM models with different sampling rates in the SisFall and FallAllD datasets are shown

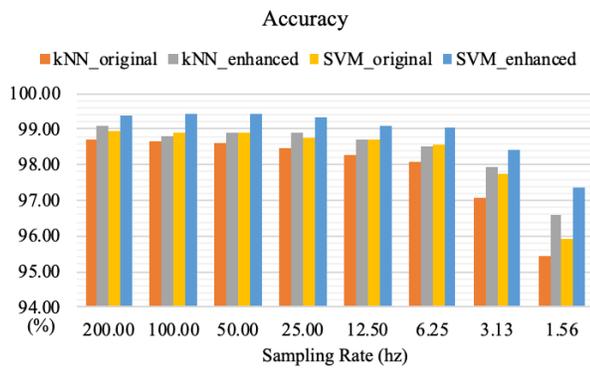

(a)

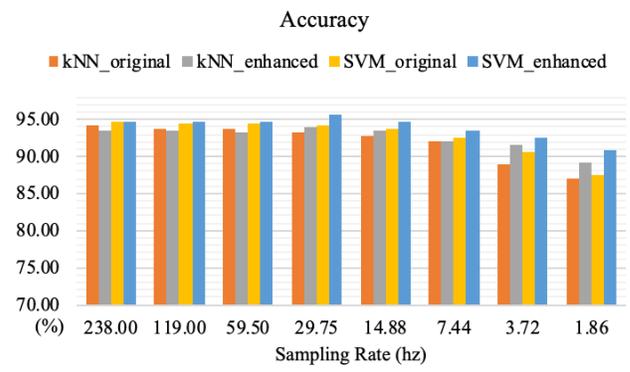

(a)

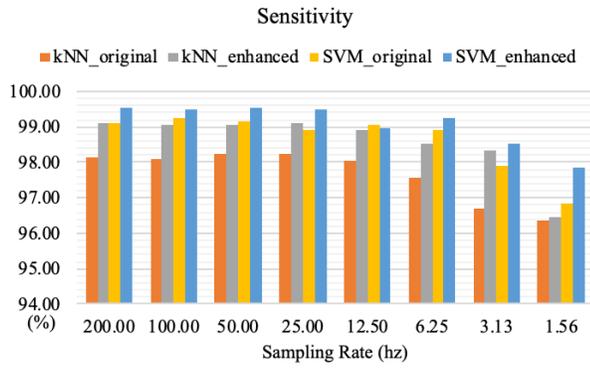

(b)

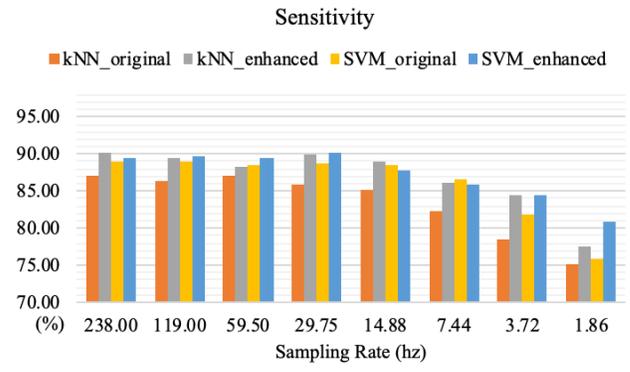

(b)

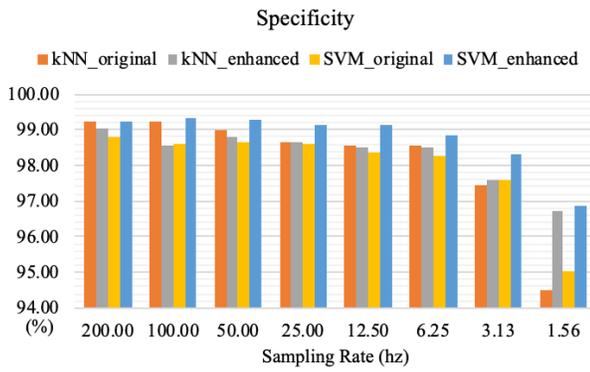

(c)

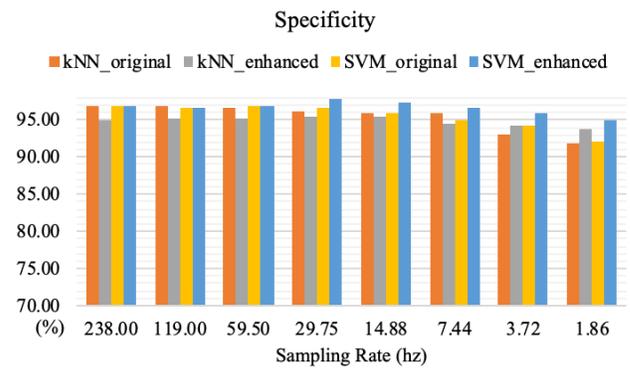

(c)

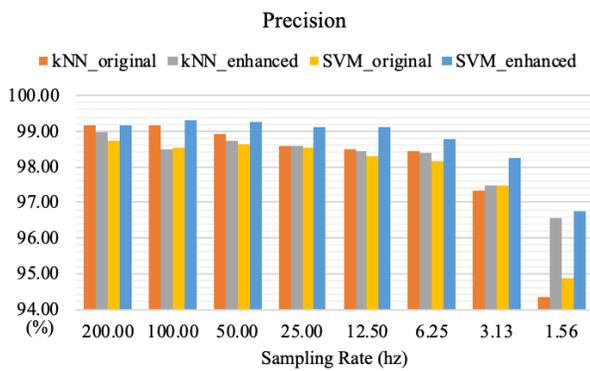

(d)

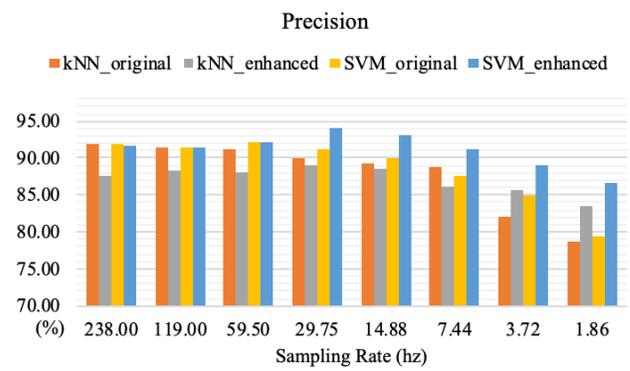

(d)

Fig. 4 (a) ACC, (b) SEN, (c) SPE, and (d) PRE of sampling rates vs typical wearable FD systems with ASE and without ASE in SisFall dataset.

Fig. 5 (a) ACC, (b) SEN, (c) SPE, and (d) PRE of sampling rates vs typical wearable FD systems with ASE and without ASE in FallAllD dataset.

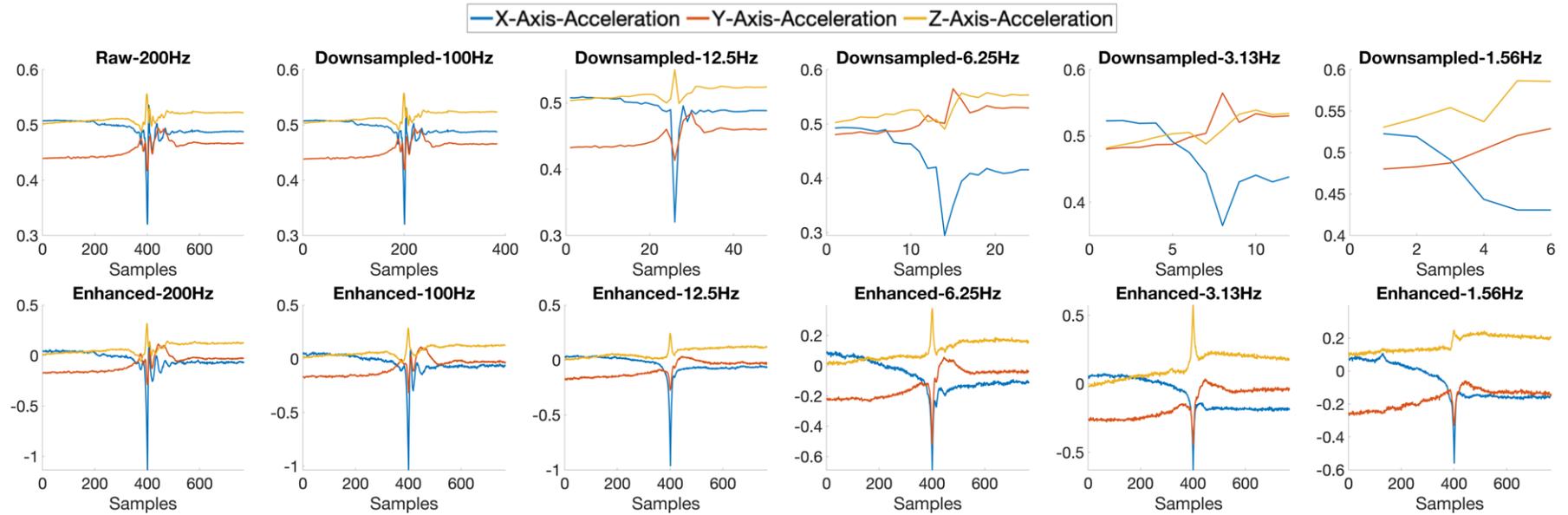

Fig. 6 Example signals of the ASE model on F14 "Fall backward while sitting, caused by fainting or falling asleep" in SisFall dataset. The 1st row shows the raw accelerometer signals and the LR signals with different downsampled sizes. The 2nd row shows the reconstructed signals using the proposed ASE model.

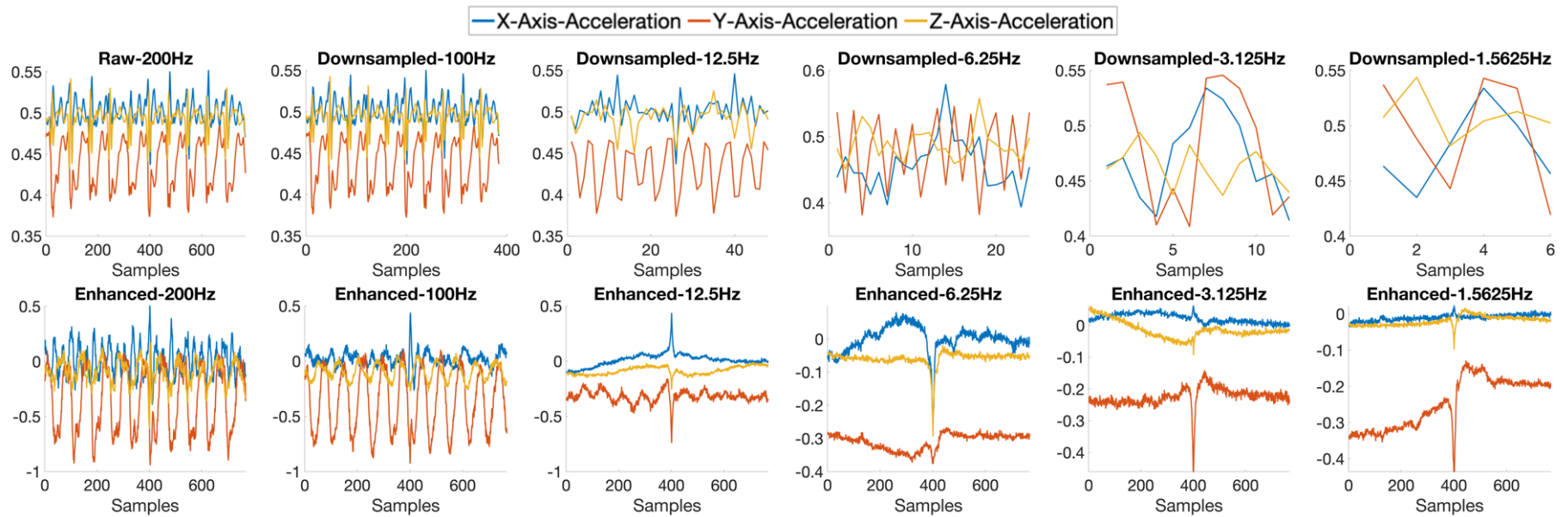

Fig. 7 Example signals of the ASE model on A03 "Jogging slowly" in SisFall dataset. The 1st row shows the raw accelerometer signals and the LR signals with different downsampled sizes. The 2nd row shows the reconstructed signals using the proposed ASE model

TABLE V
AN ANALYSIS OF POWER CONSUMPTION AND RESPONSE TIME ON ASE
MODEL IN DIFFERENT SAMPLING RATES

| Sampling Rate (Hz) | | MFLOPs | Power Consumption (mAh) | Battery Life (hour) | Response Time (sec) |
|---|---|---|---|---|---|
| FallAllD | SisFall | | | | |
| 200 | 238.00 | 915 | 1173.5 | 0.9 | 80.2 |
| 100 | 119.00 | 456 | 585.2 | 1.7 | 40.0 |
| 50 | 59.50 | 227 | 291.1 | 3.4 | 19.9 |
| 25 | 29.75 | 112 | 144.0 | 6.9 | 9.8 |
| 12.5 | 14.88 | 54.9 | 70.5 | 14.2 | 4.8 |
| 6.25 | 7.44 | 26.3 | 33.7 | 29.7 | 2.3 |
| 3.13 | 3.72 | 11.9 | 15.3 | 65.3 | 1.0 |
| 1.56 | 1.86 | 4.8 | 6.1 | 163.1 | 0.4 |

in Fig. 4 and Fig. 5, where "$k$NN_original" is the typical FD system using the $k$NN model without the proposed ASE model, and "$k$NN_enhanced" denote the $k$NN-based FD systems without and with ASE, respectively. The "SVM_original" and "SVM_enhanced" denote the SVM-based FD systems without and with ASE, respectively. Generally, applying the ASE model to FD systems improves the system performance in most sampling rates and evaluation metrics, especially accuracy and sensitivity. Moreover, the proposed ASE model has more positive effects when the sampling rate is lower. Particularly, the FD systems using SVM and ASE models outperform systems with other systems at most sampling rates. However, the performance of "$k$NN_enhanced" is lower than that of "$k$NN_original" while the sampling rate is 119 Hz and 59.5 Hz. The results show that the SVM-based FD systems with ASE models at the lowest sampling rate achieved 97.34% ACC in SisFall and 90.52% ACC in FallAllD, while that without ASE models only achieved 95.92% ACC in SisFall and 87.47% ACC in FallAllD.

The best detection results of the FD systems with and without ASE are listed in TABLE III and TABLE IV, respectively. As evident from the two tables, the best detection performance is slightly improved using the proposed ASE model, especially for the SVM models. The best accuracy of the FD systems using the ASE model achieved 99.41% for SisFall and 95.6% for FallAllD.

### B. Analysis of power consumption and response time on the ASE model

According to [33], calculating floating-point operations (FLOPs) is a common approach to estimate the computational complexity of the deep learning model. We also test the FLOPs of the proposed ASE model as indictors of the required power consumption and response time. The simulation environment comprises of the STM32L476JGY MCU with an operating frequency of 80 MHz and a 1000 mA mercury battery, and its power consumption calculator are utilized as the simulated environments. The real-world performance of FD systems depends on the accelerometer and wireless transportation modules. The FLOPs, response time, and time complexity of the proposed ASE for different sampling rates and datasets are shown in TABLE V. The proposed ASE model at the lowest sampling rate can operate for 163.1 h with a response time of 0.4 s, whereas at the original sampling rate, it only has a 0.9-h working period and a response time exceeding

1 min.

### C. Comparison with regularization approaches of ASE model for FD

Two regularization approaches involving L2 with a weight of 0.0001 and $Dropout$=0.2 are applied to test the performance of the proposed ASE on different FD approaches (e.g., $k$NN and SVM). The experimental results with the SisFall and FallAllD datasets are shown in TABLE VI and TABLE VII.

From VI and VII, the FD system using the original ASE achieves the best accuracy for SVM at most sampling rates in both datasets, whereas that using ASE with regularization (L2&Dropout) yields better performance at sampling rates of 238, 119, and 58.5 Hz in the FallAllD dataset. The results indicate that the regularization approaches work well only for signals of higher sampling rates. Similar trends are also observed in the detection performance using the ASE and ASE (L2&Dropout) models for $k$NN.

### D. Comparison with DL-based FD systems

We further verify the effectiveness of the ASE system on several DL-based FD systems, including multi-layer perception (MLP) autoencoder (AE), convolutional neural networks (CNN) AE, long short-term memory (LSTM) AE, CNN-LSTM, and deep belief neural network (DBN). The parameters employed are briefly introduced as follows.

- **MLP-AE** [34]: The model is a three-layer stacked AE model, consisting of 64, 32, and 64 neurons. The first two layers is trained to perform feature extraction, and the final layer detects fall events.
- **CNN-AE** [34]: The model is similar to MLP-AE. First, we train a two-layer 1D-CNN AE with 64 filters and a kernel size of 3 to perform feature extraction. Then, a dense layer (64 units) is adopted to detect fall events.
- **LSTM-AE** [34]: The model adopts a two-layer LSTM AE with 64 units for feature learning. Then, a dense layer (64 units) is used to detect fall events.
- **CNN-LSTM** [35]: The model is a combination of CNN and LTSM models. It consists of two 1D-CNN layers (filters = 2 and stride = 3) and one LSTM layer. A max-pooling layer is added after the first two CNN layers with a size of 2. The LSTM (64 units) and a dense layer (64 units) are then used to carry out fall event detection.
- **DBN** [36]: Unsupervised pre-training is first applied with a three-layer restricted Boltzmann machine (RBM). The hidden units of the RBM are 32, 64, and 64. A combination of the RBM and a dense layer (64 units) is then trained for FD.

The detection results using these deep learning models with and without ASE models are shown in TABLE VIII and TABLE IX, respectively. Generally, the proposed ASE model can enhance the detection performance of models at most sampling rates. However, the accuracy of several detection models using ASE in the FallAllD dataset is still slightly lower (<0.5%) than the model without ASE, including, MLP-AE at 3.72 Hz and 1.86 Hz, and CNN-LSTM at 3.72 Hz. The FD system using CNN-ASE at 238 Hz and CNN-LSTM at 14.88

TABLE VI
ACCURACY VS REGULATION METHODS IN SISFALL (%)

| ML-based Classifier | Front-end Processer | Sampling Rate (HZ) | | | | | | | |
|---|---|---|---|---|---|---|---|---|---|
| | | 200 | 100 | 50 | 25 | 12.5 | 6.25 | 3.13 | 1.56 |
| kNN | Original | 98.70 | 98.67 | 98.61 | 98.45 | 98.30 | 98.08 | 97.09 | 95.42 |
| | ASE | 99.07 | 98.79 | 98.92 | 98.89 | 98.70 | 98.52 | 97.96 | 96.60 |
| | ASE (L2) | 97.65 | 98.18 | 97.62 | 98.30 | 97.80 | 98.58 | 96.35 | 94.62 |
| | ASE (L2 & Dropout) | 98.67 | 98.61 | 98.70 | 98.52 | 98.08 | 97.90 | 96.41 | 95.30 |
| SVM | Original | 98.95 | 98.92 | 98.92 | 98.76 | 98.70 | 98.58 | 97.74 | 95.92 |
| | ASE | **99.38** | **99.41** | **99.41** | **99.32** | **99.07** | **99.04** | **98.42** | **97.34** |
| | ASE (L2) | 99.23 | 99.10 | 99.13 | 98.79 | 98.67 | 98.08 | 96.91 | 95.95 |
| | ASE (L2 & Dropout) | 99.01 | 99.10 | 98.98 | 99.10 | 98.36 | 98.98 | 97.19 | 95.70 |

TABLE VII
ACCURACY VS REGULATION METHODS IN FALLALLD (%)

| ML-based Classifier | Front-end Processer | Sampling Rate (HZ) | | | | | | | |
|---|---|---|---|---|---|---|---|---|---|
| | | 238 | 119 | 59.5 | 29.75 | 14.88 | 7.44 | 3.72 | 1.86 |
| kNN | Original | 94.11 | 93.77 | 93.83 | 93.22 | 92.82 | 91.94 | 88.96 | 87.06 |
| | ASE | 93.56 | 93.56 | 93.22 | 93.97 | 93.50 | 92.01 | 91.46 | 89.16 |
| | ASE (L2) | 93.36 | 93.83 | 93.02 | 92.41 | 91.12 | 89.30 | 88.75 | 77.85 |
| | ASE (L2 & Dropout) | 94.24 | 94.72 | 94.72 | 93.56 | 89.91 | 87.80 | 87.26 | 79.61 |
| SVM | Original | 94.58 | 94.44 | 94.51 | 94.31 | 93.83 | 92.62 | 90.65 | 87.47 |
| | ASE | 94.65 | 94.65 | 94.78 | **95.60** | 94.65 | 93.56 | **92.55** | **90.92** |
| | ASE (L2) | 94.72 | 94.78 | 94.38 | 93.56 | 92.62 | 91.19 | 90.51 | 78.12 |
| | ASE (L2 & Dropout) | **95.66** | **96.00** | **95.80** | 95.12 | 90.04 | 88.62 | 89.57 | 81.17 |

TABLE VIII
ACCURACY VS DIFFERENT DL APPROACHES IN SISFALL (%)

| ML-based Classifier | Front-end Processer | Sampling Rate (HZ) | | | | | | | |
|---|---|---|---|---|---|---|---|---|---|
| | | 200 | 100 | 50 | 25 | 12.5 | 6.25 | 3.13 | 1.56 |
| MLP-AE | Original | 98.08 | 98.18 | 97.77 | 97.90 | 97.80 | 96.44 | 96.26 | 94.93 |
| | ASE | 99.26 | 99.23 | 99.26 | 99.17 | 99.10 | 98.39 | 98.02 | 95.76 |
| CNN-AE | Original | 98.18 | 98.08 | 98.42 | 98.55 | 98.21 | 97.87 | 97.40 | 95.98 |
| | ASE | 99.10 | 99.29 | 99.29 | 99.32 | 99.13 | 98.67 | 98.18 | 95.95 |
| LSTM-AE | Original | 93.51 | 93.63 | 95.83 | 95.21 | 95.92 | 95.11 | 93.82 | 93.04 |
| | ASE | 98.92 | 99.01 | 98.86 | 99.13 | 98.92 | 98.86 | 97.93 | 96.20 |
| CNN-LSTM | Original | 95.61 | 95.27 | 96.29 | 94.56 | 95.98 | 96.13 | 94.47 | 93.38 |
| | ASE | 99.32 | 99.26 | 99.10 | 99.26 | **99.38** | 98.45 | 97.77 | 95.52 |
| DBN | Original | 98.95 | 98.76 | 97.59 | 97.06 | 96.23 | 97.50 | 96.04 | 93.63 |
| | ASE | 97.56 | 97.80 | 97.84 | 98.48 | 98.36 | 98.39 | 97.40 | 95.05 |
| kNN | Original | 98.70 | 98.67 | 98.61 | 98.45 | 98.30 | 98.08 | 97.09 | 95.42 |
| | ASE | 99.07 | 98.79 | 98.92 | 98.89 | 98.70 | 98.52 | 97.96 | 96.60 |
| SVM | Original | 98.95 | 98.92 | 98.92 | 98.76 | 98.70 | 98.58 | 97.74 | 95.92 |
| | ASE | **99.38** | **99.41** | **99.41** | **99.32** | 99.07 | **99.04** | **98.42** | **97.34** |

TABLE IX
ACCURACY VS DIFFERENT DL APPROACHES IN FALLALLD (%)

| ML-based Classifier | Front-end Processer | Sampling Rate (HZ) | | | | | | | |
|---|---|---|---|---|---|---|---|---|---|
| | | 238 | 119 | 59.5 | 29.75 | 14.88 | 7.44 | 3.72 | 1.86 |
| MLP-AE | Original | 93.25 | 94.02 | 93.62 | 92.16 | 91.88 | 91.40 | 90.84 | 90.25 |
| | ASE | 94.84 | 94.67 | 95.27 | 93.89 | 93.58 | 92.00 | 90.75 | 87.91 |
| CNN-AE | Original | 93.04 | 93.44 | 94.07 | 93.94 | 94.09 | 93.43 | 92.18 | 88.86 |
| | ASE | **95.58** | 94.88 | **95.30** | 94.56 | 94.01 | 93.43 | 92.18 | 90.92 |
| LSTM-AE | Original | 92.12 | 92.01 | 92.54 | 91.67 | 92.41 | 90.17 | 89.96 | 86.93 |
| | ASE | 94.29 | **94.93** | 95.10 | 94.77 | 93.66 | 92.04 | 91.67 | 89.79 |
| CNN-LSTM | Original | 91.95 | 92.30 | 93.13 | 91.29 | 91.27 | 91.33 | 91.24 | 87.51 |
| | ASE | 93.85 | 94.63 | 94.20 | 94.57 | 93.68 | 92.94 | 90.90 | 89.84 |
| DBN | Original | 87.61 | 88.20 | 85.85 | 85.10 | 83.02 | 72.65 | 72.65 | 72.65 |
| | ASE | 90.69 | 92.04 | 91.20 | 90.83 | 91.29 | 90.30 | 90.67 | 89.24 |
| kNN | Original | 94.11 | 93.77 | 93.83 | 93.22 | 92.82 | 91.94 | 88.96 | 87.06 |
| | ASE | 93.56 | 93.56 | 93.22 | 93.97 | 93.50 | 92.01 | 91.46 | 89.16 |
| SVM | Original | 94.58 | 94.44 | 94.51 | 94.31 | 93.83 | 92.62 | 90.65 | 87.47 |
| | ASE | 94.65 | 94.65 | 94.78 | **95.60** | 94.65 | 93.56 | 92.55 | 90.92 |

Hz can achieve the best accuracy in the FallAllD (95.60%) and SisFall (99.38%) datasets, respectively.

## V. DISCUSSION

The results reveal that the accuracy of the typical FD systems without the proposed ASE model decreases significantly when the sampling rate is less than 6.25 Hz in SisFall and 14.88 Hz in FallAllD. These results are similar to those of previous studies [19, 20]. This reduction in accuracy is owed to the loss of important information and critical movement patterns in LR signals. As shown in Fig. 6 and Fig. 7, the accelerometer signals degrade significantly when the sampling rate decreases from the initial sampling rate (Raw-200 Hz) to an extremely low sampling rate (Downsampled-1.56 Hz). Moreover, it leads to several technical challenges for FD systems such as mismatch of the effective features and degradation effects.

To deal with these challenges, an ASE model is proposed to reconstruct the accelerometer signals from LR to HR. As presented in Fig. 6, several important movement features of the fall signals (Enhanced-1.56 Hz) are reconstructed using the proposed ASE model such as impact, free-fall, and vibration. The reconstructed signals can help FD systems tackle degradation effects and the mismatch problems of the effective features. Moreover, the proposed ASE model has positive effects on HR-FD systems, as shown in TABLE III and TABLE IV. It filters the noise (e.g., muscle vibration) and keep the motion patterns, which improves the ability of the FD system to classify falls and ADL.

Fig. 7 demonstrates that the enhanced ADL patterns are more similar to a fall than an ADL, especially the sampling rate is lower than 6.25 Hz, where the reconstructed signals lack periodicity. However, the enhanced signals still enable the FD system to achieve better detection performance. This is because the proposed ASE model strengthens the differences between the ADL and fall patterns compared to the raw patterns. The FD system has the potential to obtain more effective features from the enhanced signals to distinguish falls from ADL. Future work is to develop more powerful signal reconstruction models for long-term telehealthcare monitoring systems. Additional techniques will be applied to reconstruct the movement characterizes, e.g., periodicity.

The proposed ASE model can enhance the detection performance of several well-known DL methods, including MLP-AE, CNN-AE, LSTM-AE, CNN-LSTM and DBN. However, the best performance of DL methods is slightly worse than that of typical ML methods (e.g., kNN and SVM). Therefore, further refinement of the combination of DL-based FD and ASE model is required in future studies. Moreover, we plan to design a lightweight model for power efficiency.

The experimental results show that FD systems using SVM and ASE models can achieve the best accuracy with the SisFall and FallAllD datasets. Similarly, various fall systems [37, 38] have shown the superiority of SVM for FD systems owing to its good generalization capacity. However, SVM still has several technical challenges, including large-scaling problems, online SVM training, kernel function selection and parameter optimization [39]. We plan to design a more advanced mechanism to support FD systems to tackle these challenges as these issues are critical during the development of SVM-based FD systems. Furthermore, cloud-based parameter optimization based on [40] may improve efficiency and performance.

To the best of our knowledge, this is the first study focusing on improving the performance of LR-FD systems using the DL-based ASE model. A similar signal enhancement approach has been successfully implemented in other fields such as face recognition [41], speech recognition [42] and enhancement [43, 44]. For a typical FD system, the proposed ASE model could be a front-end processor to enhance the LR signals before further processing. The results demonstrate that the ASE model enables LR-FD systems to achieve better detection accuracy. Additionally, the extra computational cost of the ASE model is acceptable and valuable for FD systems as the power consumption of the data sampling is much greater than any other process, even using ML or DL models [23, 24].

Previous studies have applied other power-saving techniques to FD systems that achieved over 90% detection performance [17, 18]. However, it is difficult to directly compare the design principle as the objective of this study is different from early studies. Firstly, they proposed a power-efficient FD method that can adjust or optimize sampling rates to balance the detection performance and sampling rate selection. Different from these previous works, we focus on directly enhancing the LR-FD system using the DL-based ASE model. Hence, the proposed ASE model has great potential to be incorporated into their studies to achieve better system performance when dealing with LR accelerometer signals. We plan to fuse the proposed ASE model and [17, 18] to develop energy-efficient FD systems in the future. Secondly, the energy-efficient mechanisms of [17, 18] were validated on private datasets. It is difficult to assess the effects of the sampling rates and effectiveness of the proposed ASE model on their datasets. This is because the impact of sampling rates on different datasets is diverse. For example, in this study, the sampling rate has a greater impact on the public FallAllD dataset than on the public SisFall dataset when more complicated experiments are employed. Finally, the proposed ASE model is developed for ML-based FD systems. However, the power-efficient mechanisms [17, 18] are designed for rule-based detection FD algorithms instead of ML-based FD algorithms, making it difficult to apply their approaches to ML-based FD algorithms directly.

To investigate the potential feasibility of the proposed ASE model in real-time implementation, we have estimate the response time and power consumption of the proposed ASE model based on FLOPs [33]. Considering the feasibility and usability in the real world, the proposed ASE model at 25 Hz with a working period of 6.9 h and 9.8 s is more suitable for real-world implementation. However, further research on designing a lightweight ASE model is required for real-time detection and mobile devices. In several real-world FD systems [12], the sensor side only capture and transmit signals to processing units. The proposed ASE model is implemented in the server side for signal enhancement and FD because of its powerful computational resources.

The main limitation of this study is that the applied open datasets involve emulated fall events and ADL performed by healthy young subjects rather than real-world data from elder people. Their movement characteristics and features may be different from those of the emulated experiments [45]. This can lead to the model, trained with the emulated data, performing worse on real-world data [46]. Therefore, a real-world dataset (e.g., FARSEEING [47]) will be included to validate the effectiveness of the proposed ASE model on FD systems in the future.

## VI. CONCLUSION

To enhance the performance of LR-FD systems, technical challenges such as mismatch of effective features and degradation effects, are required for further investigation. In this study, we propose a DL-based ASE model that reconstructs the HR signals from the LR sampling by learning the relationship and correlation between the LR and HR signals. The predicted fine-grained movement information from the enhanced accelerometer signals enables LR-FD systems to tackle technical challenges and achieve better performance. The results show that the ASE model can efficiently help LR-FD systems achieve better system performance.

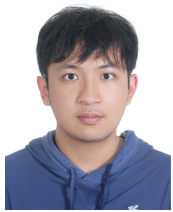

**Kai-Chun Liu** received the M.S. and Ph.D. degree in biomedical engineering from National Yang-Ming University, Taipei, Taiwan, in 2015 and 2019. He is currently a Postdoctoral Scholar with the Research Center for Information Technology Innovation, Academia Sinica, Taipei. His research interests include pervasive healthcare, wearable computing, machine learning and biosignal processing.

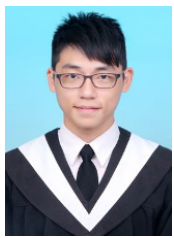

**Kuo-Hsuan Hung** received the B.S. and M.S. degrees from the National Chiao Tung University and National Central University, Taiwan, in 2015 and 2017, respectively. He is currently working toward the Ph.D. degree with the Department of Biomedical Engineering, National Taiwan University. He is a Research Assistant with the Research Center for Information Technology Innovation, Academia Sinica, Taipei, Taiwan. His research interests include biomedical signal processing, noise reduction, speaker recognition, and deep learning.

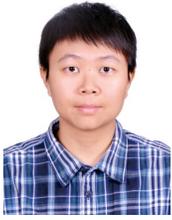

**Chia-Yeh Hsieh** received the Ph.D. degree in biomedical engineering from National Yang-Ming University, Taipei, Taiwan, in 2019. From December 2019 to July 2021, she was a Postdoctoral Research Fellow with the Department of Biomedical Engineering, National Yang-Ming University, Taipei, Taiwan. Since August 2021, she has been an Assistant Professor with the Bachelor's Program in Medical Informatics and Innovative Applications, Fu Jen Catholic University, New Taipei City, Taiwan.

Her research interests include exercise-cognition interaction, fall detection, activity recognition, indoor positioning, rehabilitation assessment and health monitoring systems.

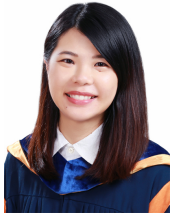

**Hsiang-Yun Huang** received the M.S. degree in biomedical engineering from National Yang-Ming University, Taipei, Taiwan, in 2015, and Ph.D. degree in biomedical engineering from National Yang Ming Chiao Tung University, Taipei, Taiwan, in 2021. Her research interests include indoor positioning, activity recognition, fall detection, rehabilitation assessment and health monitoring systems.

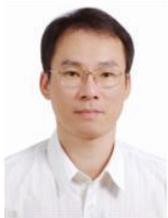

**Chia-Tai Chan** received the Ph.D. degree in computer science and information engineering from National Chiao Tung University, Hsinchu, Taiwan, in 1998.

From 1999 to 2005, he was a Project Researcher with Telecommunication Laboratories Chunghwa Telecom Co., Ltd. In August 2005, he joined the faculty of the Institute of Biomedical Engineering, National Yang-Ming University, as an Associate Professor. He is currently a professor in the Department of Biomedical Engineering, National Yang Ming Chiao Tung University. His research interests include Active and Assisted Living: Technologies and Applications, Personalized Health Information Technologies.

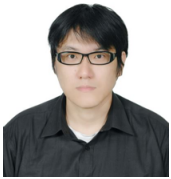

**Yu Tsao** (Senior Member, IEEE) received the B.S. and M.S. degrees in electrical engineering from National Taiwan University, Taipei, Taiwan, in 1999 and 2001, respectively, and the Ph.D. degree in electrical and computer engineering from the Georgia Institute of Technology, Atlanta, GA, USA, in 2008. From 2009 to 2011, he was a Researcher with the National Institute of Information and Communications Technology, Tokyo, Japan, where he engaged in research and product development in automatic speech recognition for multilingual speech-to-speech translation. He is currently a Research Fellow (Professor) and Deputy Director with the Research Center for Information Technology Innovation, Academia Sinica, Taipei, Taiwan. He is also a Jointly Appointed Professor with the Department of Electrical Engineering at Chung Yuan Christian University, Taoyuan City, Taiwan. His research interests include assistive oral communication technologies, audio coding, and bio-signal processing. He is currently an Associate Editor for the IEEE/ACM TRANSACTIONS ON AUDIO, SPEECH, AND LANGUAGE PROCESSING and IEEE SIGNAL PROCESSING LETTERS. He received the Academia Sinica Career Development Award in 2017, National Innovation Awards in 2018, 2019, and 2020, Future Tech Breakthrough Award 2019, and Outstanding Elite Award, Chung Hwa Rotary Educational Foundation 2019–2020.